\documentclass[]{spie}  

\pdfoutput=1

\def\apjl{ApJ}
\def\apj{ApJ}

\def\aap{Astron. \& Astrophys.}

\def\jcap{JCAP}
\def\procspie{Proc. SPIE}
\def\ao{Applied Optics}
\def\pasj{PASJ}

\usepackage{url}
\usepackage{amsmath,amsfonts,amssymb}
\usepackage{graphicx}
\usepackage[colorlinks=true, allcolors=blue]{hyperref}

\title{Multimode bolometer development for the PIXIE instrument}

\author[a,b]{Peter C. Nagler}
\author[c]{Kevin T. Crowley}
\author[a]{Kevin L. Denis}
\author[a]{Archana M. Devasia}
\author[d]{Dale J. Fixsen}
\author[d]{Alan J. Kogut}
\author[a]{George Manos}
\author[e]{Scott Porter}
\author[a]{Thomas R. Stevenson}
\affil[a]{Code 553, NASA/Goddard Space Flight Center, Greenbelt, MD 20771 USA}
\affil[b]{Department of Physics, Brown University, Providence, RI 02912 USA}
\affil[c]{Department of Physics, Princeton University, Princeton, NJ USA}
\affil[d]{Code 665, NASA/Goddard Space Flight Center, Greenbelt, MD 20771 USA}
\affil[e]{Code 662, NASA/Goddard Space Flight Center, Greenbelt, MD 20771 USA}

\authorinfo{Further author information: Send correspondence to Peter C. Nagler\\E-mail: peter.c.nagler@nasa.gov, Telephone: 1-301-286-5983}

\begin{document} 
\maketitle

\begin{abstract}
The Primordial Inflation Explorer (PIXIE) is an Explorer-class mission concept designed to measure the polarization and absolute intensity of the cosmic microwave background. In the following, we report on the design, fabrication, and performance of the multimode polarization-sensitive bolometers for PIXIE, which are based on silicon thermistors. In particular we focus on several recent advances in the detector design, including the implementation of a scheme to greatly raise the frequencies of the internal vibrational modes of the large-area, low-mass optical absorber structure consisting of a grid of micromachined, ion-implanted silicon wires. With $\sim30$ times the absorbing area of the spider-web bolometers used by Planck, the tensioning scheme enables the PIXIE bolometers to be robust in the vibrational and acoustic environment at launch of the space mission. More generally, it could be used to reduce microphonic sensitivity in other types of low temperature detectors. We also report on the performance of the PIXIE bolometers in a dark cryogenic environment.
\end{abstract}


\section{INTRODUCTION}\label{sec:intro}

The Primordial Inflation Explorer (PIXIE)\cite{Kogutetal2011,Kogutetal2014} is a space-based polarizing Fourier transform spectrometer (FTS) designed to measure the polarization and intensity spectra of the cosmic microwave background (CMB). As for previous FTS-based instruments flown to measure the CMB \cite{WoodyandRichards1981, Gushetal1990, Matheretal1994, Tuckeretal1997}, PIXIE's design and experimental approach represent a significant departure from the focal plane imagers most commonly used for these measurements. This is especially true for the detectors. Instead of requiring several thousand diffraction-limited, ultra-low-noise detectors, PIXIE can achieve nK-scale sensitivity across 2.5 decades in frequency with just four multimode polarization-sensitive bolometers based on silicon thermistors. With a large etendue $A\Omega$ of $4$ cm$^{2}$ sr per detector, the detectors are designed for a high optical load ($\sim 120$ pW), but their noise equivalent power (NEP) is near the thermodynamic limit and is subdominant to photon noise from the CMB. Using high impedance silicon thermistor-based bolometers allows the use of simple and mature junction field effect transistor (JFET)-based voltage amplifiers. As a nulling experiment, where the signal is a small modulated component in a bright background, the detectors will always operate where assumptions of linearity are strong. A series of discrete symmetries built into the instrument enable multiple detectors to measure the same signal, allowing measurement and control of detector-sourced systematic effects. Developing detectors for a FTS with a large but mechanically robust absorbing area ($\sim30$ times that of Planck's spider-web bolometers \cite{Holmesetal2008}), large enough bandwidth and appropriate geometry to measure optical frequencies from 15 GHz to 5 THz, and sufficiently low NEP ($\le 1\times10^{-16}$ W$\sqrt\mathrm{{Hz}}$) requires meeting a unique set of design, fabrication, and performance criteria. We describe these in this paper.

\section{INSTRUMENT DESCRIPTION}\label{sec:inst}

\begin{figure}
\centering
\includegraphics[width=0.33\textwidth]{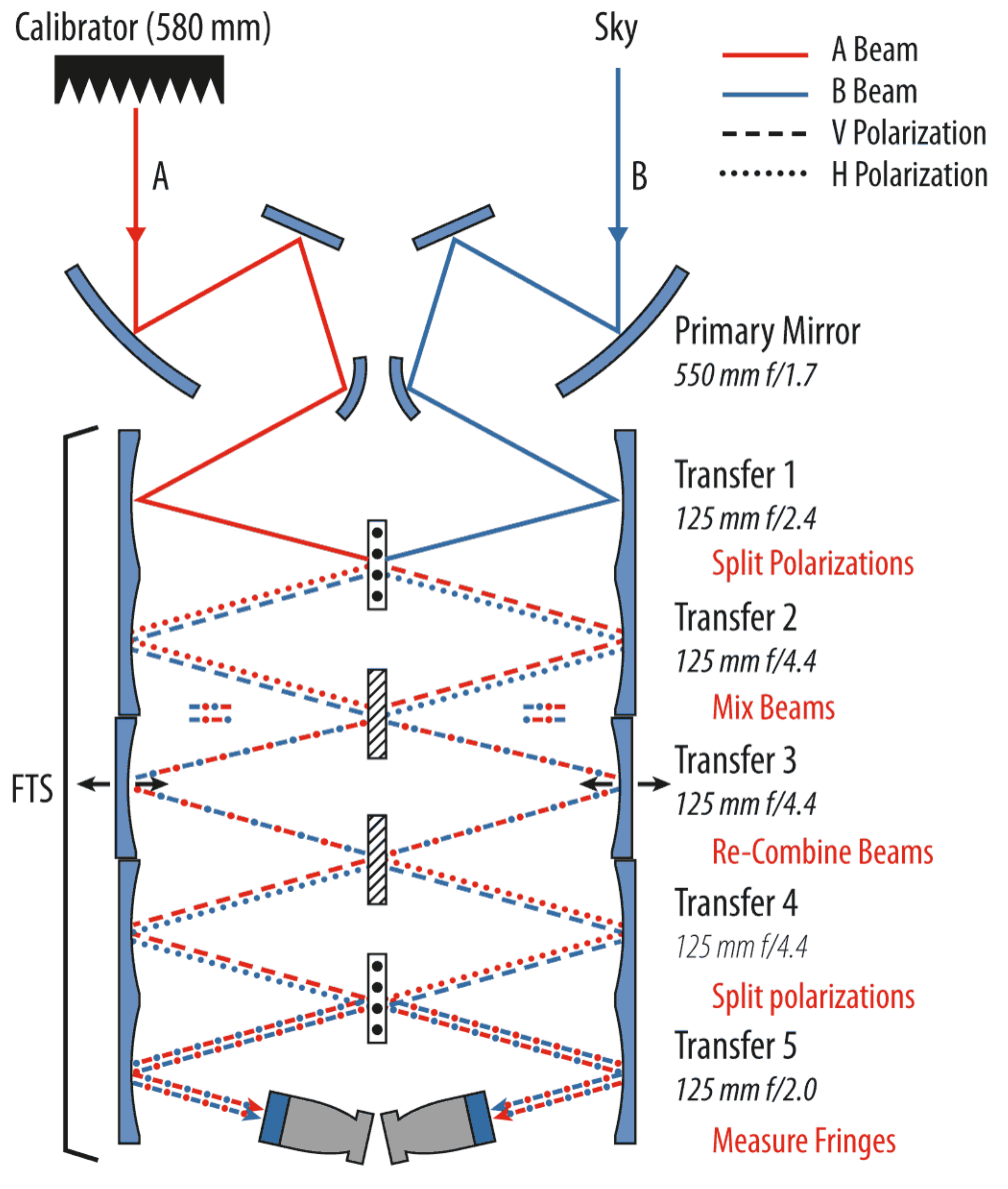}
\raisebox{0.10\height}{\includegraphics[width=0.45\textwidth]{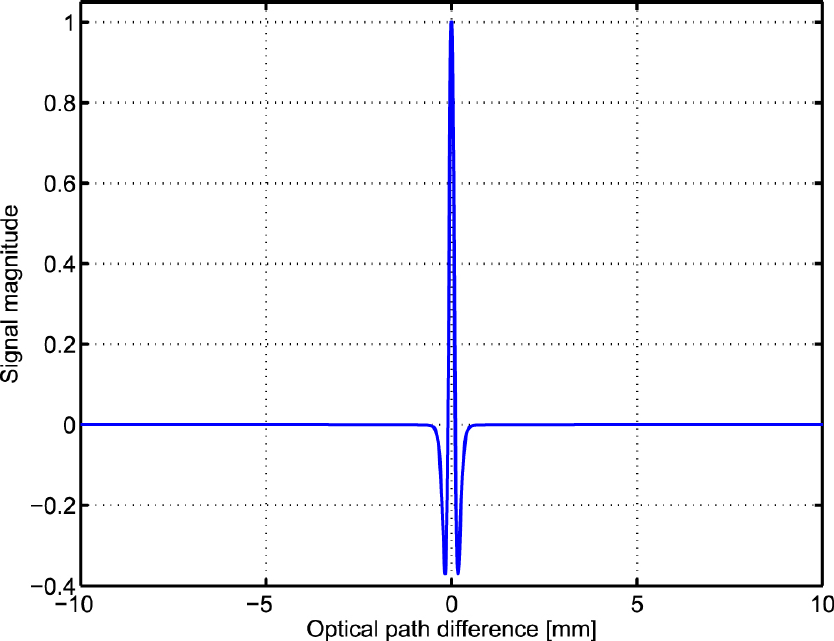}}
\caption{Left: Cartoon of the PIXIE FTS. The listed dimentions give the diameters of components. The calibrator can block either beam or be stowed. The polarizers are free standing wire grids, reflecting one polarization and transmitting the other. The moving mirror's position $z$ is $\pm2.6$ mm, which corresponds to an optical path difference $\ell$ of $\pm10.0$ mm. The mirror completes a stroke from $z=-2.6$ mm to $z=+2.6$ mm in 3 seconds. The optical path difference between beams $\ell$ is related to the frequency of incident radiation $\nu$ by the relationship $\ell=c/\nu$, where $c$ is the speed of light. The frequency of the mirror movement $\omega$ is related to the frequency of incident radiation by $\omega=4\nu v/c$. where $v$ is the moving mirror's velocity. The CMB signal is largely confined to acoustic frequencies below 15 Hz. The dust signal is confined to acoustic frequencies below 100 Hz. These constraints drive the bolometer bias and bandwidth requirements. Right: Simulated time domain signal (interferogram) incident on a PIXIE bolometer. This shows the Fourier transform of the polarized CMB.}
\label{fig:FTS}
\end{figure}

Complete descriptions of the PIXIE instrument are available\cite{Kogutetal2011,Kogutetal2014}, but here we reproduce the highlights that drive bolometer design and performance requirements. A cartoon of the PIXIE FTS is shown in Figure \ref{fig:FTS}. Light is directed into the FTS by the primary mirrors, folding flats, and secondary mirrors. Polarizer A defines the instrument's polarization basis, transmitting horizontal ($\hat{x}$) polarization and reflecting vertical ($\hat{y}$) polarization. Polarizer B is oriented at 45$^{\circ}$ relative to polarizer A and mixes the beams. The moving mirror injects an optical phase delay. Polarizer C has same orientation as polarizer B and sorts the beams. Polarizer D has the same orientation as polarizer A and again splits polarizations. Light is then directed into polarization-maintaining receiver horns\cite{Kogutetal2015} and onto the focal planes. Each focal plane consists of two bolometers mounted back-to-back with their polarization axes orthogonal, allowing a simultaneous measurement of both linear polarization states. The low frequency cutoff of the instrument ($\sim 15$ GHz) is set by the etendue, and the high frequency cutoff ($\sim 5$ THz) is set by a series of filters and the polarizer grid spacing.

Light incident on the instrument is represented by $\vec{E}_{inc} = A\hat{x}+B\hat{y}$. If both beams are open to the sky, the power measured by the detectors is

\begin{equation}\label{eq:P}
\begin{aligned}
{\bf P}_{x}^{L}& = \frac{1}{2}\int\big(A^{2}+B^{2}\big)+\big(A^{2}-B^{2})\cos\Big(\frac{4\nu z}{c}\Big)d\nu,\\
{\bf P}_{y}^{L}& = \frac{1}{2}\int\big(A^{2}+B^{2}\big)+\big(B^{2}-A^{2})\cos\Big(\frac{4\nu z}{c}\Big)d\nu,\\
{\bf P}_{x}^{R}& = \frac{1}{2}\int\big(A^{2}+B^{2}\big)+\big(A^{2}-B^{2})\cos\Big(\frac{4\nu z}{c}\Big)d\nu,\\
{\bf P}_{y}^{R}& = \frac{1}{2}\int\big(A^{2}+B^{2}\big)+\big(B^{2}-A^{2})\cos\Big(\frac{4\nu z}{c}\Big)d\nu,
\end{aligned}
\end{equation}
where the superscripts $L$ and $R$ indicate whether the detector is on the left of right side of the FTS (see Figure \ref{fig:FTS}), the subscripts $x$ and $y$ indicate whether the detector is measuring $\hat{x}$ or $\hat{y}$ polarization, $\nu$ is the frequency of incident radiation, $z$ is the mirror position, and $c$ is the speed of light.

Each power expression given in Equation \ref{eq:P} consists of a DC term and a term modulated by the movement of the mirror. The former represents the total intensity of incident light, and the latter is the Fourier transform of the difference spectrum between orthogonal incident polarizations (Stokes $Q$ in instrument-fixed coordinates). Given the small linearly polarized fraction of the microwave sky, the signal measured by the detectors thus consists of a small modulated component on top of near-constant optical bias. Taking the inverse Fourier transform of the measured signals removes the DC component and we are left with the spectra of polarized light measured by each detector:

\begin{equation}\label{eq:S}
\begin{aligned}
{\bf S}_{x}^{L}\left(\nu\right)& = A_{\nu}^{2}-B_{\nu}^{2},\\
{\bf S}_{y}^{L}\left(\nu\right)& = B_{\nu}^{2}-A_{\nu}^{2},\\
{\bf S}_{x}^{R}\left(\nu\right)& = A_{\nu}^{2}-B_{\nu}^{2},\\
{\bf S}_{y}^{R}\left(\nu\right)& = B_{\nu}^{2}-A_{\nu}^{2},
\end{aligned}
\end{equation}
where the subscript $\nu$ indicates that we are working in the frequency domain. The PIXIE bolometers are designed to measure this signal.

\section{DETECTOR DESIGN AND FABRICATION}

\subsection{Overview}

\begin{figure}[t]
\centering
\includegraphics[width=0.45\textwidth]{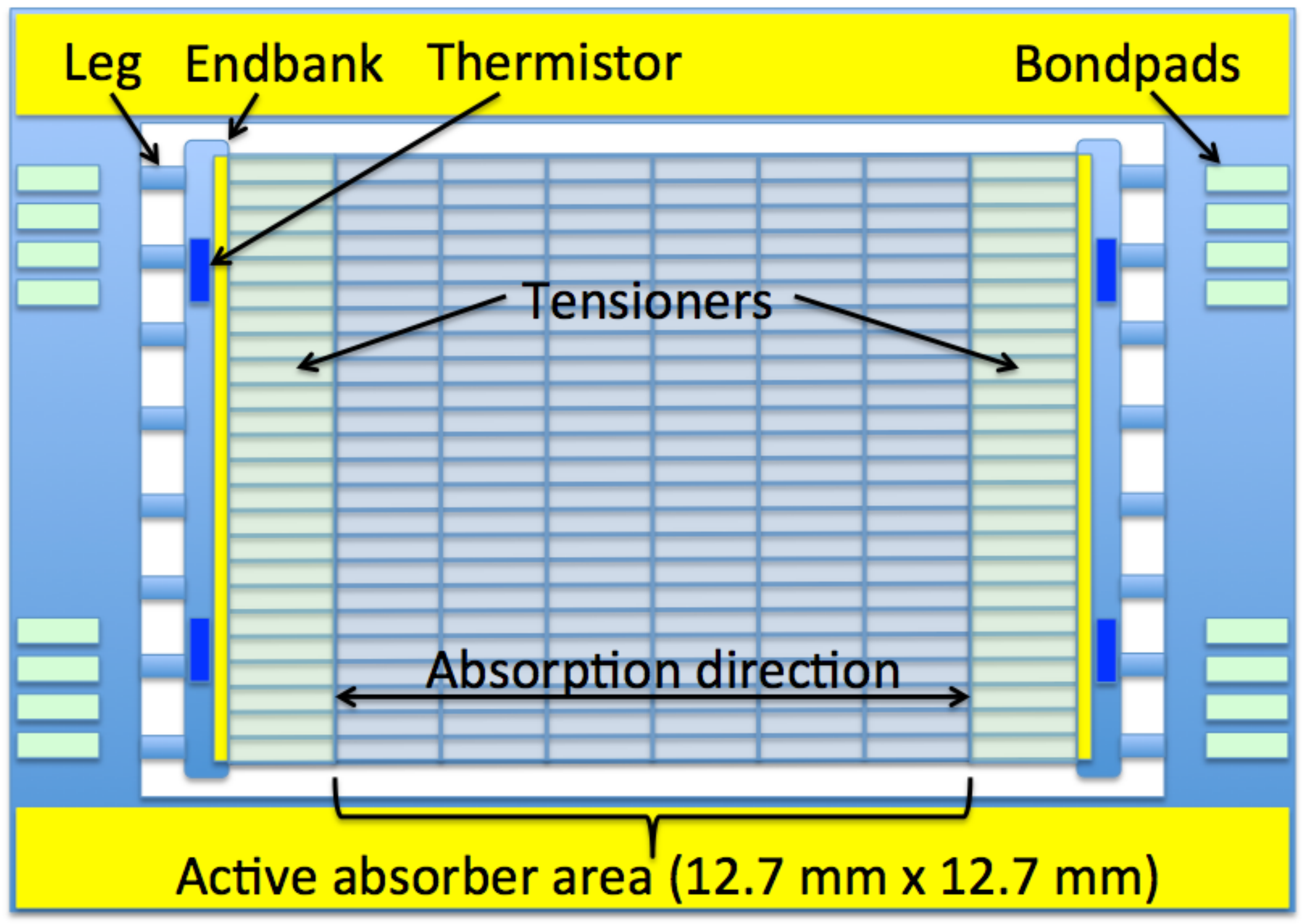}
\includegraphics[width=0.52\textwidth]{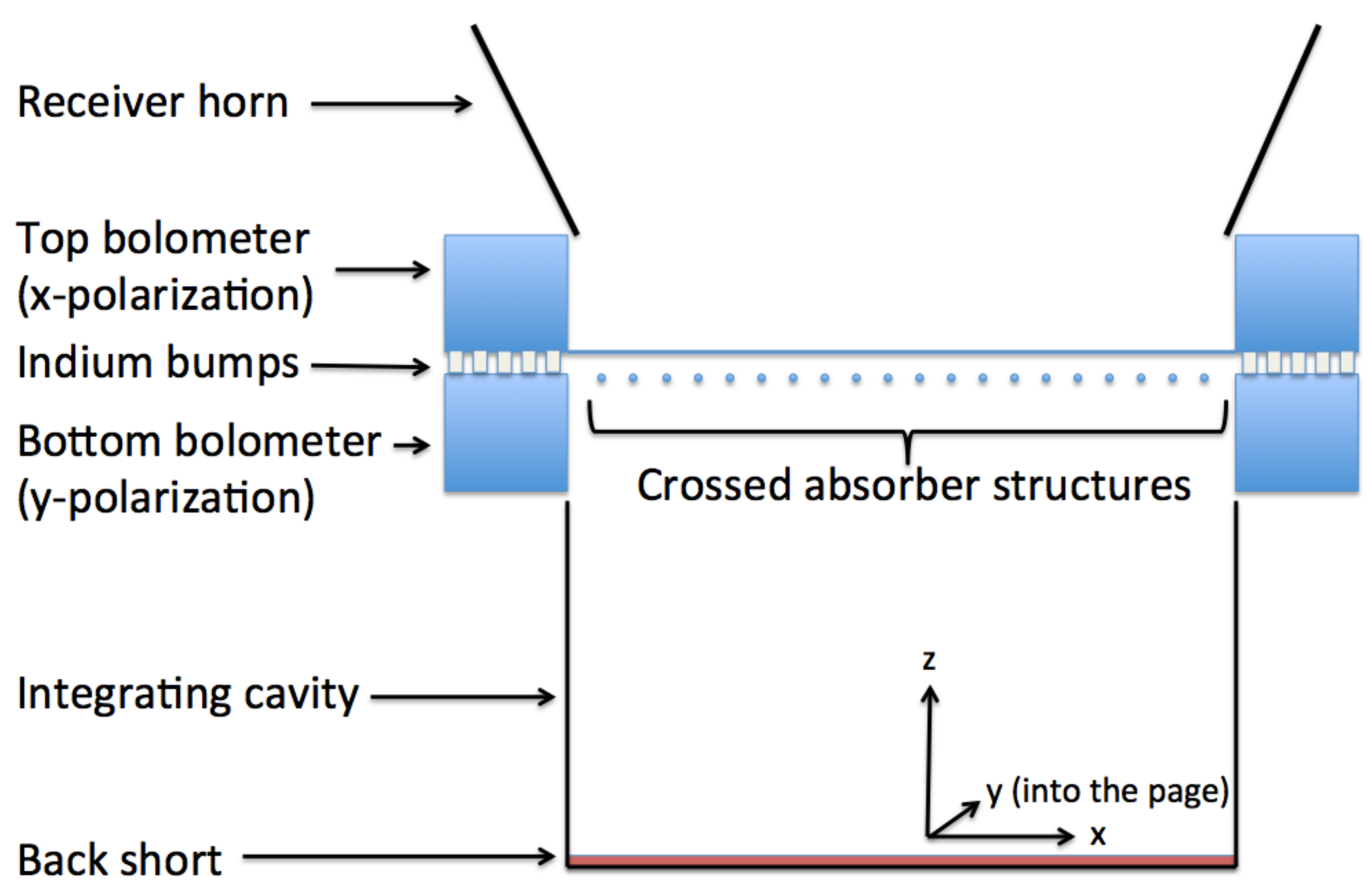}
\caption{Cartoon of a PIXIE bolometer, showing an entire chip (left) and a profile view of a hybridized pair of bolometers (right). Degenerately doped wires of single-crystal silicon absorb a single linear polarization along the indicated direction of absorption. Undoped support beams orthogonal to the direction of light absorption provide mechanical support while contributing a negligible cross-polar response. The response time of the detector is driven by the heat capacity $C$ of the endbank's gold thermalization bar and the effective thermal conductance $G_{e}$ (which accounts for electrothermal feedback gain) of the silicon legs between the endbank and the chip frame. These can be independently tuned. With indium bumps the pair of chips will be stood off from each other by $<20$ $\mu$m.}
\label{fig:bolo_cart}
\end{figure}

A cartoon of a PIXIE bolometer and a cross sectional view of a hybridized pair of bolometers is shown in Figure \ref{fig:bolo_cart}. The bolometer fabrication and ion implantation are performed in the Detector Development Lab (DDL) at NASA/GSFC. The bolometer consists of an optical absorber structure coupled to silicon thermistor-based thermometers that operate at $\sim200$ mK under optical and electrical bias. The thermometers are coupled to the chip frame held at the bath temperature ($\sim 100$ mK). The absorber structure (Section \ref{sec:abs}) is a large-area (1.61 cm$^{2}$) grid of freestanding, micromachined, degenerately doped silicon wires. The thermistors are located on silicon membranes (``endbanks'', Section \ref{sec:eb}) at either end of the absorber structure. Each endbank consists of a gold bar for thermalization and two thermistors that are independently biased and read out for laboratory characterization but will be wired in series or parallel in flight. The endbanks are coupled to the chip frame (Section \ref{sec:frame}) through a series of silicon legs. On the chip frame are wirebond pads for bias and readout of the thermistors and an array of indium bumps enabling the hybridization of a pair of bolometer chips. When hybridized, orthogonal absorbers will be separated by $<20$ $\mu$m, such that the two bolometers will sample orthogonal polarizations of nearly the same electric field.

\subsection{Absorber stucture}\label{sec:abs}

\begin{figure}[t]
\centering
\includegraphics[width=.49\textwidth]{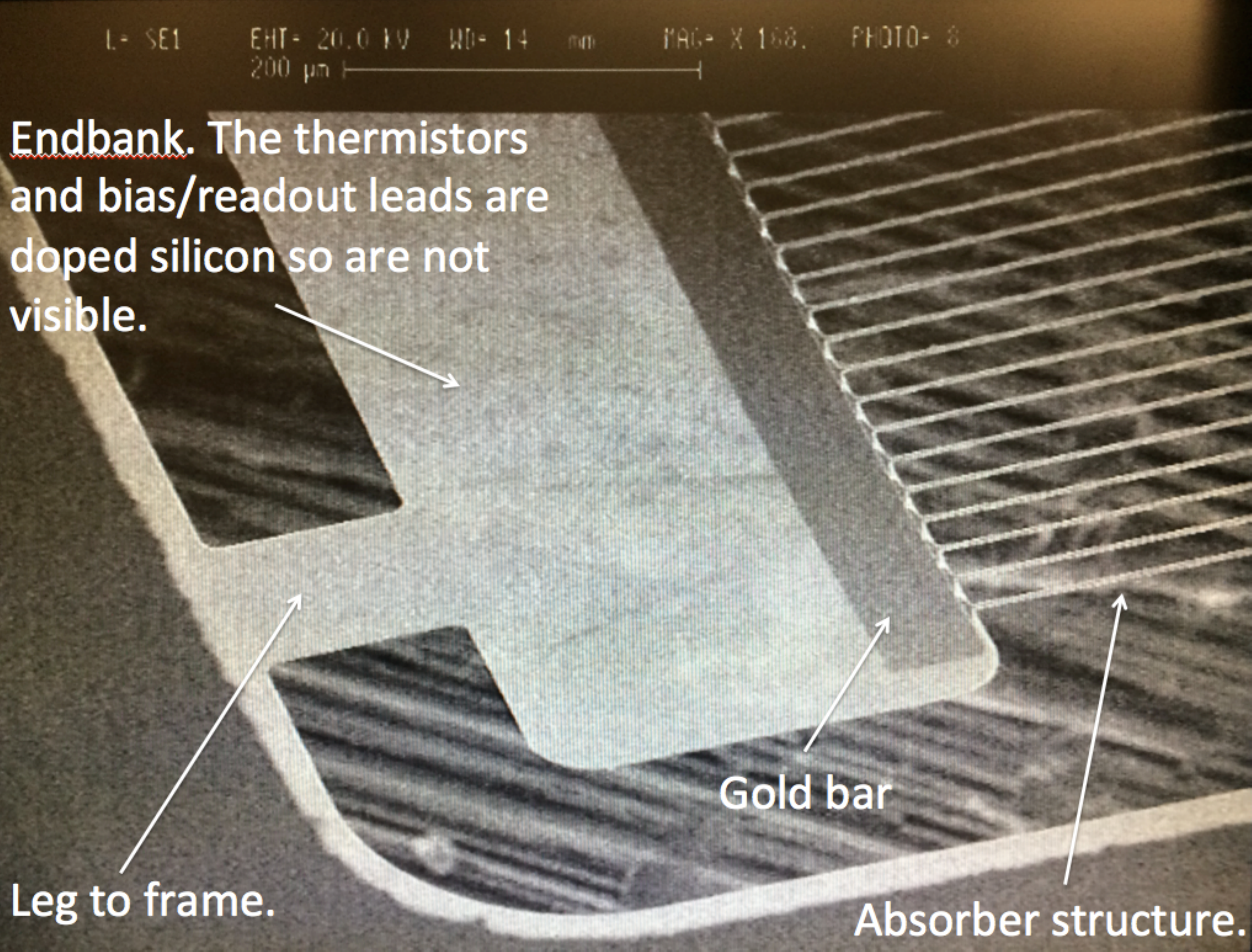}
\includegraphics[width=.5\textwidth]{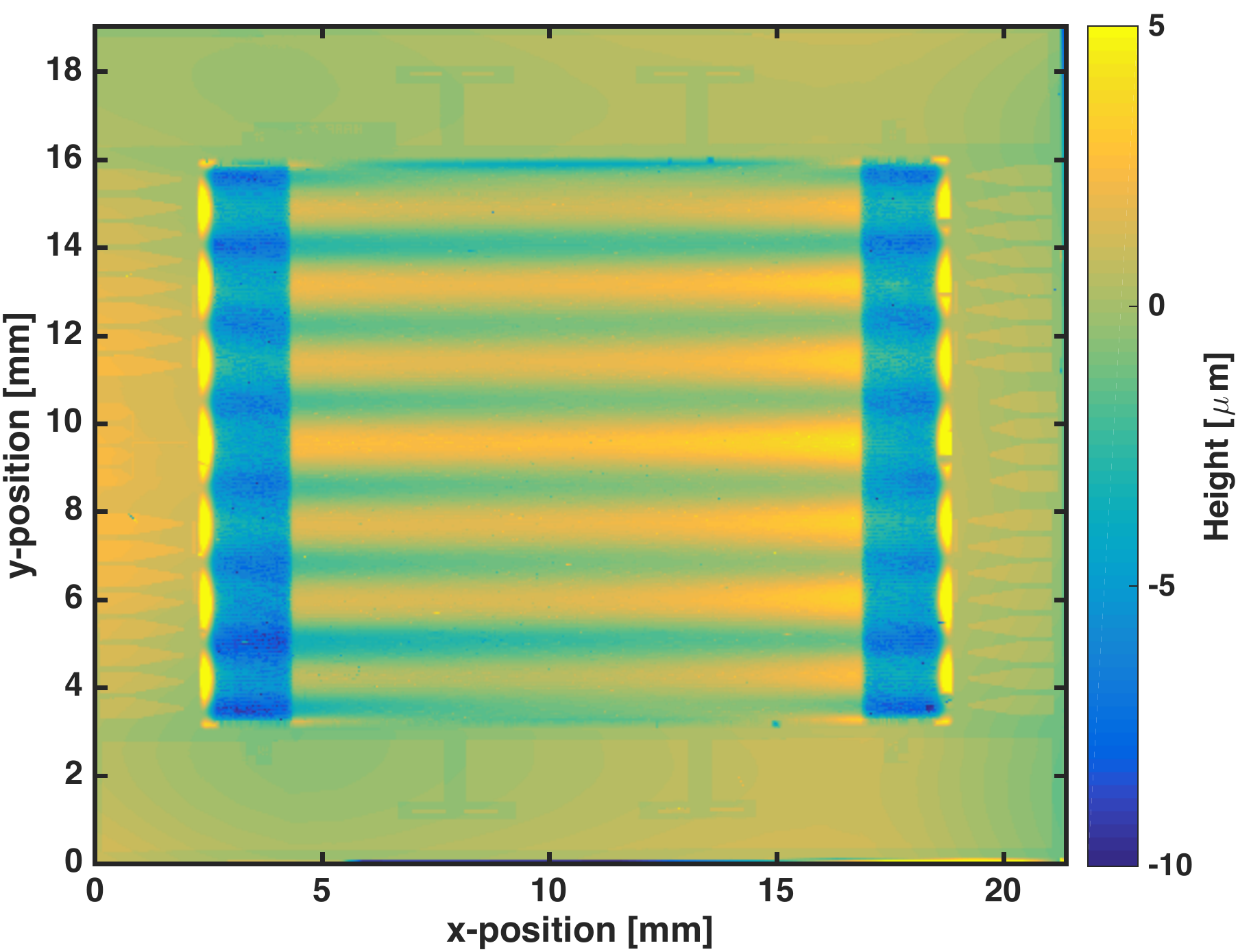}
\caption{Left: Scanning electron micrograph of a completed bolometer showing the interface between the absorber structure and the endbank. Right: White light interferometer image showing the flatness of the absorber achieved after implementing the tensioning scheme. Along the $\hat{x}$-direction (the direction of absorption), the absorbers are flat to less than 2 $\mu$m. The periodic structure visible along the $\hat{y}$-direction is set by the endbank leg placement, but with an amplitude of $<5$ $\mu$m and a period of several mm, they absorbers are essentially flat. The chip frame is visible around the edge of the image. Since the strings never protrude more than $\sim5$ $\mu$m above the frame, a hybridized pair can be offset by fewer than 20 $\mu$m without risk of strings colliding during launch.}
\label{fig:abs}
\end{figure}

Shown in Figure \ref{fig:abs}, the bolometer absorber structure consists of an array of parallel micromachined silicon wires that absorbs one linear polarization\cite{Kusakaetal2014}. They are 3 $\mu$m wide, 1.35 $\mu$m thick, 15.8 mm long, and are on a 30 $\mu$m pitch, making them effective absorbers for frequencies up to $\sim 5$ THz. The wires are degenerately doped with phosphorous to be metallic at all temperatures. The absorber structure's effective sheet resistance is designed to be 377 $\Omega/\square$, impedance matching it to free space. The string width and thickness are highly uniform across the array. The thickness is set by the device layer thickness on the starting silicon-on-insulator (SOI) substrate, and they are etched to width with an inductively-coupled plasma (ICP) reactive ion etch (RIE) process. 

In previous generations of devices, we found that the doping process induced compressive stress in absorber strings along the axis of absorption, causing individual strings to buckle and protrude $\pm 15$ $\mu$m from the chip frame. This is problematic for a hybridized pair of chips which are separated by $<20$ $\mu$m, since absorbers could collide and break, particularly during vibrations at launch. To prevent this, and to raise the resonant frequencies of the strings to well above the frequencies they will experience at launch, we deposited a highly-tensile Al$_{2}$O$_{3}$ film on individual strings outside the active absorbing area. This feature effectively pulls the strings taught (see Figure \ref{fig:abs}), and based on our elastic model should keep the amplitude of any string vibrations to less than $0.4$ $\mu$m rms over the band of excitation frequencies expected during launch (20 Hz - 10 kHz).

We also estimated the bolometer sensitivity to cosmic ray hits. To minimize the impact of cosmic ray hits, the time constant of the absorber structure should be much shorter than the hit rate. Based on Planck's measured hit rate \cite{Planck2014}, and scaling for our geometry, we expect one cosmic ray hit every two seconds. Measurements and thermal models of our absorber structure yield a time constant of a few ms, so particle hits can easily be flagged and removed during data processing without corrupting an entire interferogram. 

\subsection{Endbanks}\label{sec:eb}

Each bolometer chip has two endbanks, one at each end of the absorber structure (Figure \ref{fig:abs}). The endbanks consist of a gold bar for thermalization and two doped silicon thermistors on a crystalline silicon membrane. The gold bar also sets the heat capacity of the endbank. The endbank membrane is the same thickness (1.35 $\mu$m) as the absorber strings, and is also formed from the device layer of the SOI substrate. Endbanks are connected to the chip frame through eight silicon legs.

The thermistors are doped to operate below their metal-insulator transition. The electron transport mechanism in this regime is variable range hopping (VRH)\cite{shklovskiiandEfros2013}, which yields the following resistance-temperature relationship:

\begin{equation}\label{eq:VRH}
R\left(T\right)=R_{0}\times\exp\sqrt{\frac{T_{0}}{T}},
\end{equation}
where $R_{0}$ and $T_{0}$ are constants determined largely by the device geometry and doping concentration, respectively.

We also developed a detailed thermal model of the endbank that takes into account heat flow between each thermal element (absorber strings, gold bar, thermistor and legs). This will be discussed in greater detail in Section \ref{sec:perf}.

\subsection{Frame}\label{sec:frame}

\begin{figure}
\centering
\includegraphics[width=0.5\textwidth]{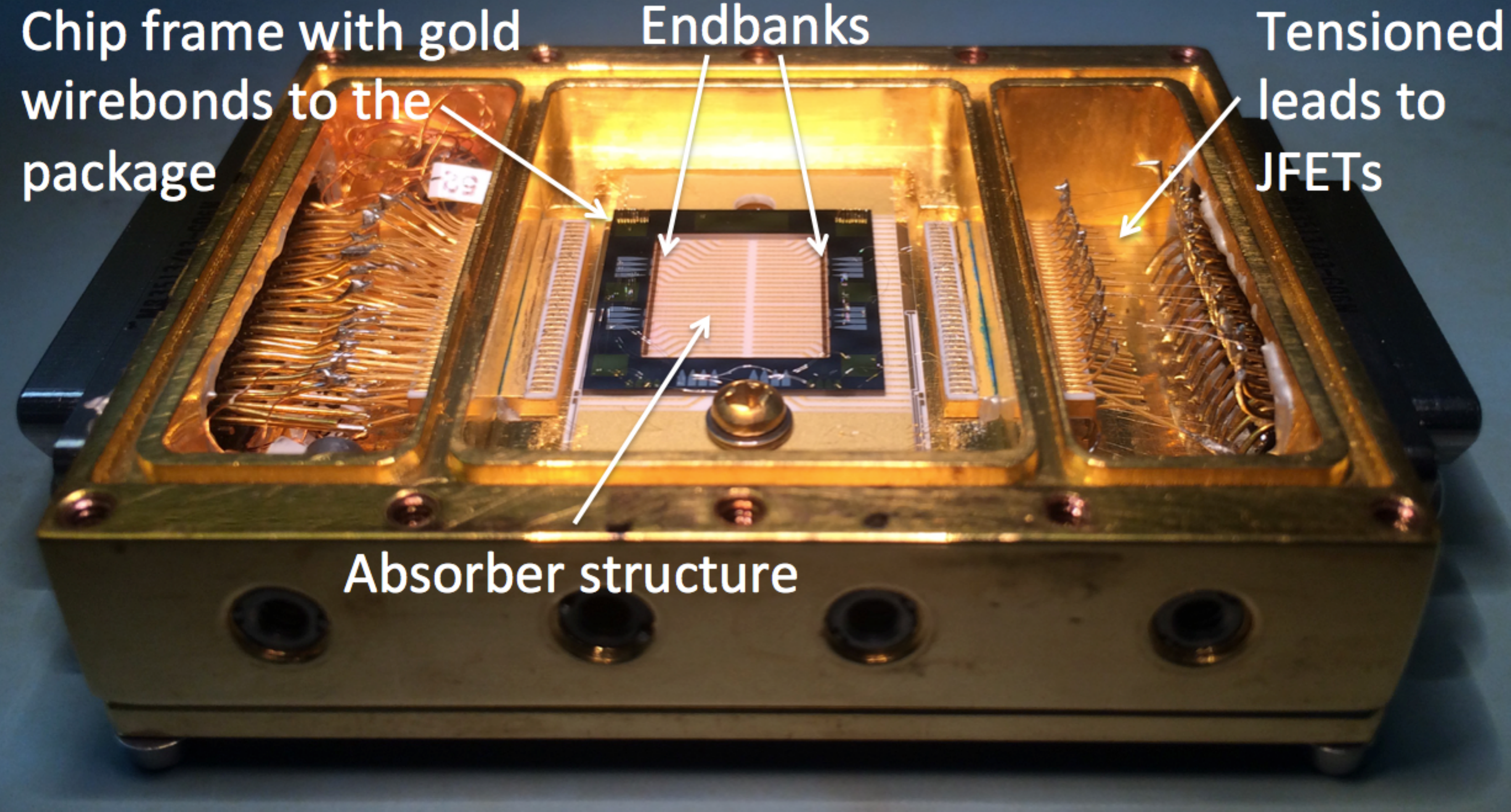}
\includegraphics[width=0.49\textwidth]{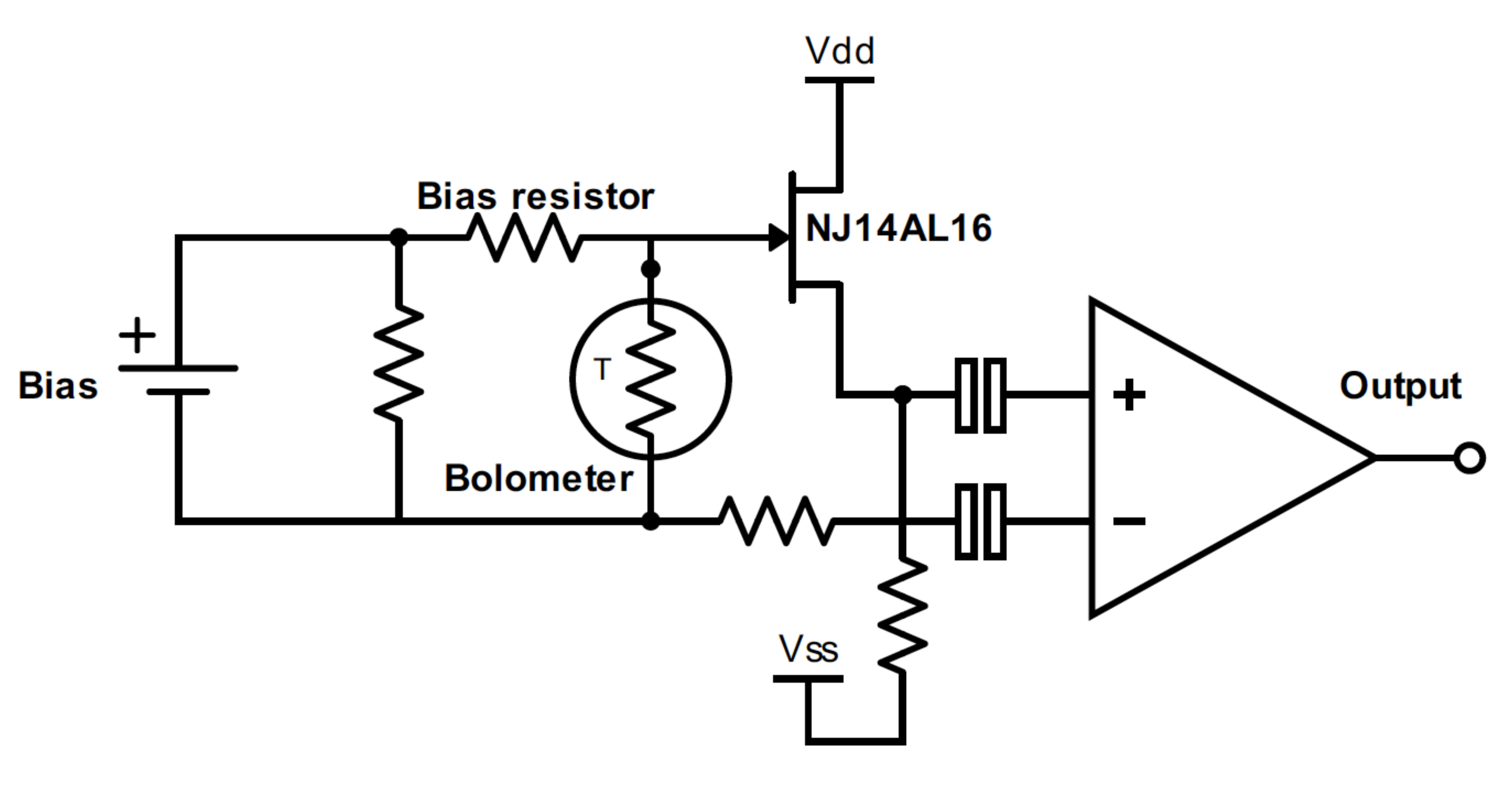}
\caption{Left: Single bolometer packaged for dark tests in the Astro-E2/Suzaku dewar. This bolometer was fabricated with $3/4$-length absorbers, enabling it to fit in a package designed for an x-ray microcalorimeter array. Right: Basic bias and readout circuit used for dark detector characterization. This is not a complete schematic, but rather a representation of what is used.}
\label{fig:packaged}
\end{figure}

The bolometer frame is maintained at the refrigerator bath temperature. Under normal operation, the bath will be kept at $\sim 100$ mK. The chip is maintained at this temperature through a series of gold wire bonds connecting gold heat sink areas on the chip to the package. The frame is designed such that any two bolometer chips can be indium bump hybridized together, though we can also characterize individual bolometers in the light or dark (Figure \ref{fig:packaged}). The frame has several arrays of indium bumps. Some are designed for mechanical purposes, and others are electrical, allowing the thermistors on one chip to be read out through wirebond pads on the other chip.

\section{DETECTOR PERFORMANCE}\label{sec:perf}

We measured detector load curves and noise in a dark cryogenic environment at a range of temperatures ($100-2000$ mK). A photograph of a packaged bolometer and a schematic of the basic bias and readout circuit are shown in Figure \ref{fig:packaged}. The package, bias and readout system, and dewar were originally built to characterize the silicon thermistor-based x-ray microcalorimeters developed for the Astro-E2/Suzaku mission \cite{Kelleyetal2007}. The thermistor is wired in series with a bias resistor. The bias resistor's resistance is much greater than the thermistor's operating resistance, so the thermistor is under current bias. Tensioned leads connect the bolometer to a cryogenic JFET amplifier, mitigating capacitive microphonic contamination of the signal band. For PIXIE, we use InterFET NJ14AL16 JFETs\cite{Interfet} that operate at $130$ K in a source-follower configuration. These are the same model JFETs that were flown as part of the XRS instrument on Astro-E2/Suzaku\cite{Kelleyetal2007} and the SXS instrument on Astro-H/Hitomi\cite{Porteretal2010}. They convert the high source impedance of the thermistors (M$\Omega$-scale) to the low output impedance of the JFETs ($1.8$ k$\Omega$). The JFETs have excellent noise performance ($5.5$ $\mathrm{nV}/\sqrt{\mathrm{Hz}}$ at 100 Hz). The low impedance signal is then AC coupled to a room temperature amplifier.

\begin{figure}
\centering
\includegraphics[width=0.49\textwidth]{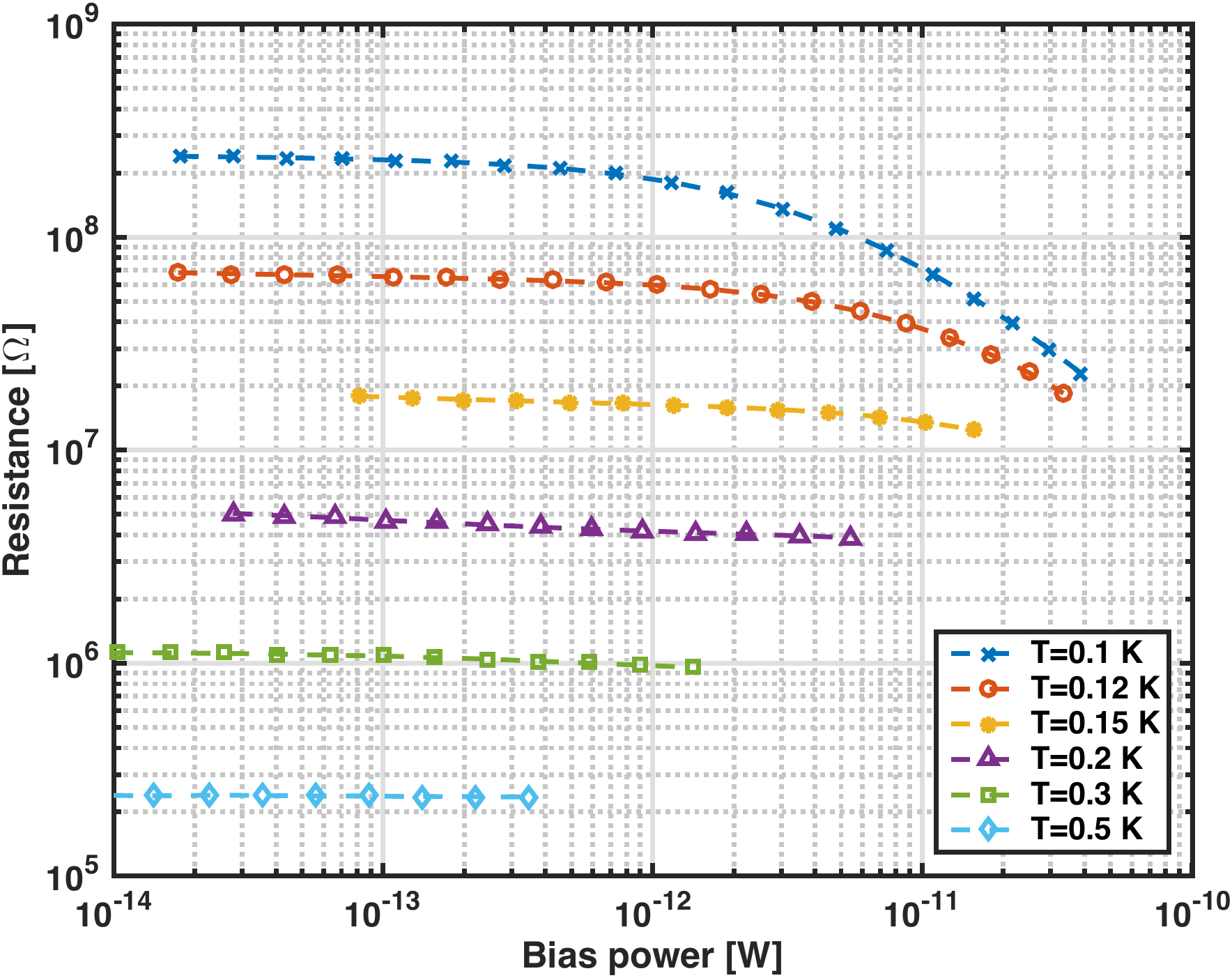}
\includegraphics[width=0.5\textwidth]{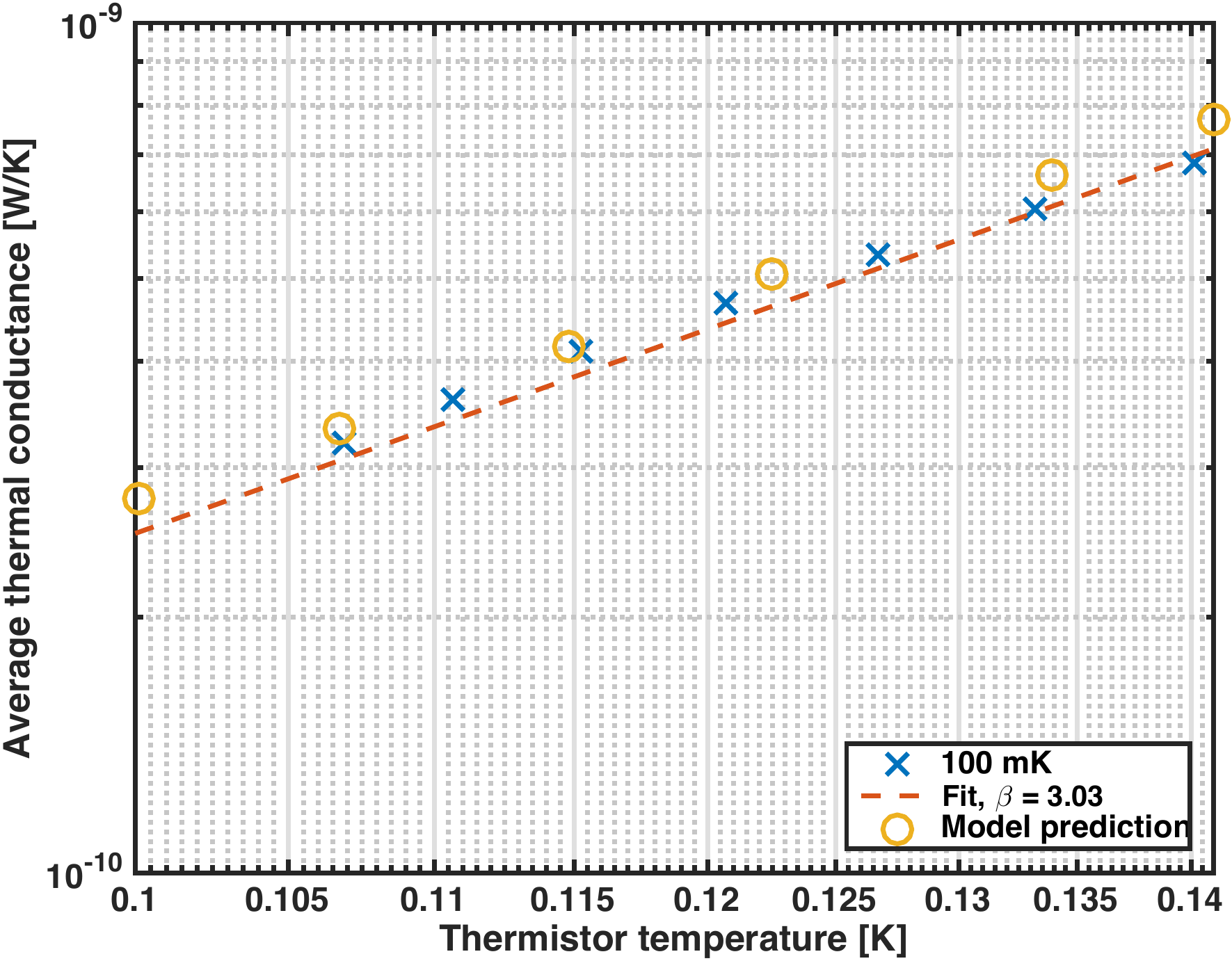}
\caption{Left: Series of measured load curves plotted to show the bolometer resistance $E/I$ as a function of bias power $I\times E$, where $E$ is the voltage drop across the thermistor and $I$ is the thermistor bias current. Data are shown for a pair of thermistors wired in series on one of the two endbanks. The data from the other endbank are similar. Because the bias circuit is optimized for low temperature, high impedance operation, it could not supply enough current for us to observe thermistor self-heating at the higher bath temperatures. From the low bias resistances, where the thermistor is isothermal with the bath, we can extract the VRH parameters. Right: Average thermal conductance $\bar{G}$ as a function of thermistor temperature. The legend shows the bath temperature corresponding to the measurement. Shown in circles is the model prediction.}
\label{fig:bias_noise}
\end{figure}

\subsection{Load curves}\label{sec:loadcurves}

A series of measured load curves is shown in Figure \ref{fig:bias_noise}, plotted to show thermistor resistance as a function of bias power. From the measured resistances under low electrical bias, where the thermistor is isothermal with the bath, we determine the VRH parameters $R_{0}$ and $T_{0}$ (Equation \ref{eq:VRH}). For this device, fitting to the data gives $T_{0}=15.11$ K and $R_{0}=911$ $\Omega$. The expected operating resistance under optical bias is $5.42$ M$\Omega$, matching well the input impedance of the JFETs. 

The average thermal conductance $\bar{G}$ between the thermistors and the bath is calculated from the high bias end of the load curves where self-heating is observed. In a steady state, the bias power $P_{\mathrm{bias}}$ is equal to the heat flow from the thermistor to the bath, so $\bar{G}$ is given by

\begin{equation}\label{eq:heat}
\bar{G}=\frac{P_{\mathrm{bias}}}{T_{1}-T_{2}},
\end{equation}
where $T_{1}$ is the thermistor temperature (calculated from Equation \ref{eq:VRH}), and $T_{2}$ is the bath temperature. This is plotted for the 100 mK data in Figure \ref{fig:bias_noise}. We then fit to the measured $\bar{G}$ with a function $\tilde{G} = G_{0}\times T^{\beta}$, also shown in Figure \ref{fig:bias_noise}. The fit is close to the expected value for phonon conduction at these temperatures ($\beta=3$).

\begin{figure}
\centering
\includegraphics[width=0.49\textwidth]{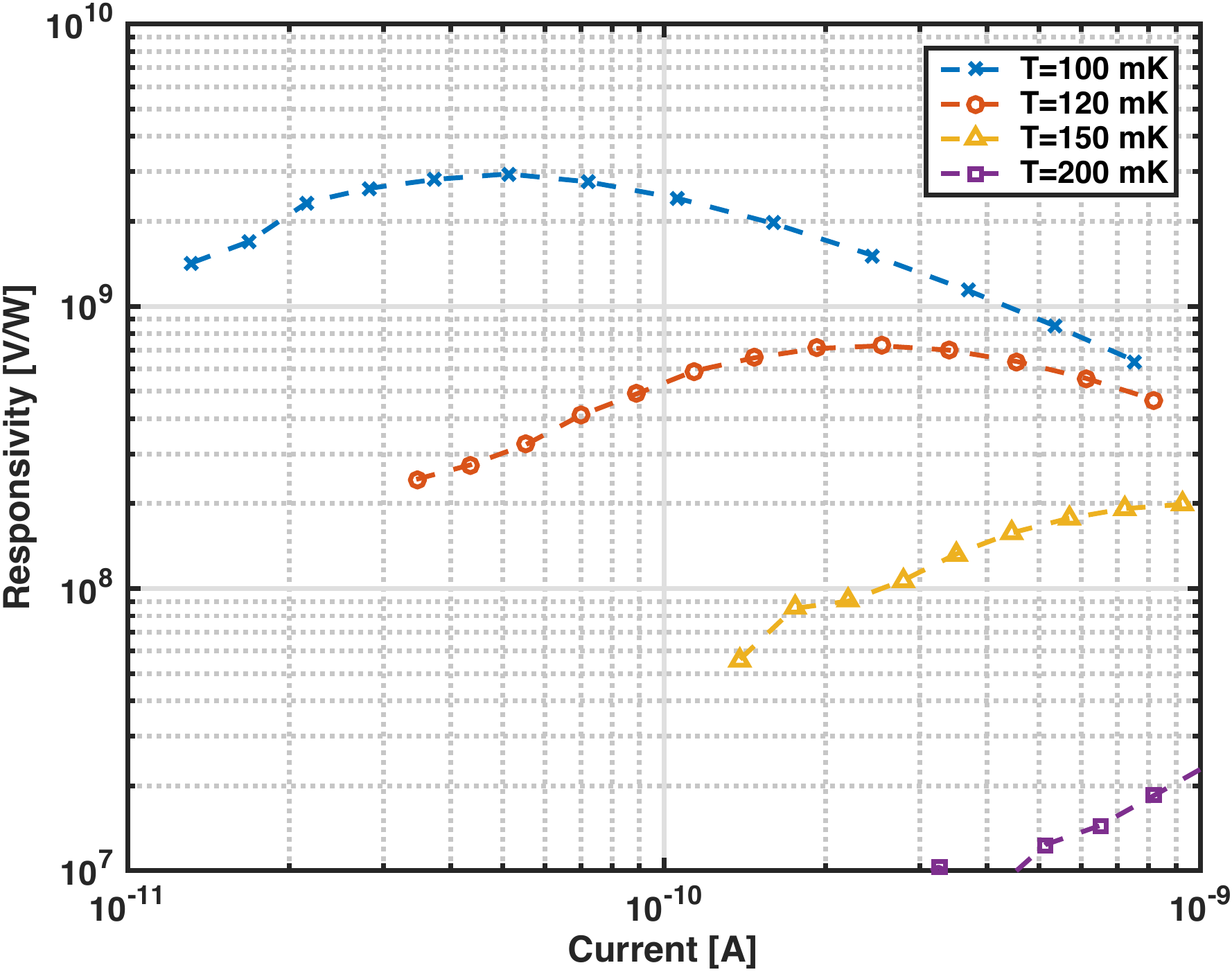}
\includegraphics[width=0.49\textwidth]{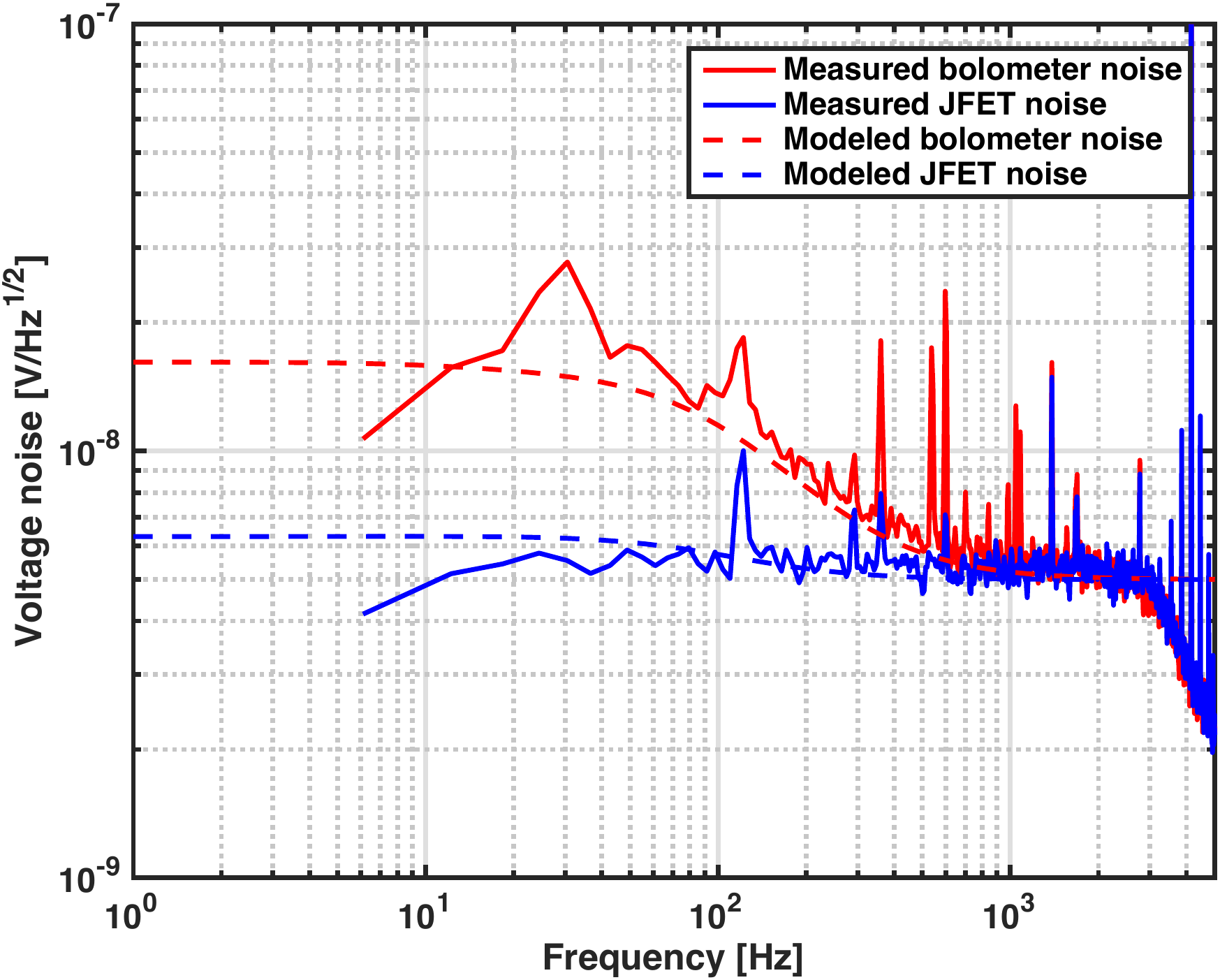}
\caption{Left: DC electrical responsivity $S_{\mathrm{e}}$ as a function of bias current for a few bath temperatures. Right: measured noise spectrum along with the fit from our thermal model. Note that the measurement was AC coupled, so both measured signals drop off at low frequency. 
Running this model for the optical and electrical bias conditions in expected in flight, we calculate a bolometer NEP of $7.93\times10^{-17}$ W/$\sqrt{\mathrm{Hz}}$. This is well below the expected photon noise from the CMB across the entire PIXIE bandwidth (see Figure \ref{fig:NEP}).}
\label{fig:Se_noise}
\end{figure}

The DC electrical responsivity $S_{\mathrm{e}}$ is also determined from the load curves. Following the notation of Jones \cite{Jones1953,Richards1994}, the responsivity is given by the following expression:

\begin{equation}\label{eq:Se}
S_{\mathrm{e}}=\frac{R-Z}{2IR},
\end{equation}
where $R$ is the resistance $E/I$ and $Z$ is is the dynamic resistance $\mathrm{d}E/\mathrm{d}I$, determined from the load curve. The measured responsivity $S_{\mathrm{e}}$ is plotted as a function of current $I$ for a few temperatures in Figure \ref{fig:Se_noise}.

We compared the measured bolometer characteristics to predictions from a thermal model that takes into account heat exchanges between the gold bar, the thermistors, and the legs. Given the endbank geometry, we solve for the etendue between each thermal component, then determine the heat flow between each of them. Heat flow between components on the endbank is via ballistic phonon conduction, where the phonon mean free path $\ell_{\mathrm{mfp}}$ is much greater than the endbank dimensions. This assumption is based on thermal transport measurements made on far infrared microwave kinetic inductance detectors (MKIDs)\cite{Brownetal2010}, where $\ell_{\mathrm{mfp}}$ in silicon at 100 mK was measured to be $\ge 1$ mm\cite{Rostem2016}. For heat flowing out the legs, we account for a mean free path reduction due to phonon scattering off the leg edges. The model is then used to predict the DC thermal conductance (Figure \ref{fig:bias_noise}) and the noise (Section \ref{sec:noise}).

\subsection{Noise analysis}\label{sec:noise}

Noise spectra were measured at a range of temperatures and bias currents. When PIXIE observes the CMB the bolometers will operate under high optical bias ($\sim 120$ pW). This results in a large temperature difference between the thermistors and the bath. It is therefore necessary to take into account non-equilibrium effects in the noise \cite{Mather1982}. 

Accounting for non-equilibrium effects, the NEP of a simple bolometer in the dark is given by

\begin{equation}\label{eq:NEP}
\mathrm{NEP_{bolometer}}^{2} = \gamma_{1}4{k_{b}}T^{2}G + \frac{1}{S^{2}}\Big(\gamma_{2}4{k_{b}}TR + e_{n}^{2}+\gamma_{3}i_{n}^{2}R + \gamma_{4}\mathrm{NEP_{excess}}^{2}\Big),
\end{equation}
where the constants $\gamma_{1}$, $\gamma_{2}$, and $\gamma_{3}$ account for non-equilibrium effects, $e_{n}^{2}$ and $i_{n}^{2}$ are the amplifier's voltage and current noise spectral densities, and $\mathrm{NEP_{excess}}$ accounts for other sources of noise, for example parasitic resistance in the leads or stray light. The factor of $1/S^{2}$, where $S$ is the bolometer's electrical responsivity, refers the Johnson, amplifier, and excess noise to the bolometer's input. Multiplying the NEP$^{2}$ by $S^{2}$ gives the voltage power spectral density of the bolometer.

\begin{figure}
\centering
\includegraphics[width=0.49\textwidth]{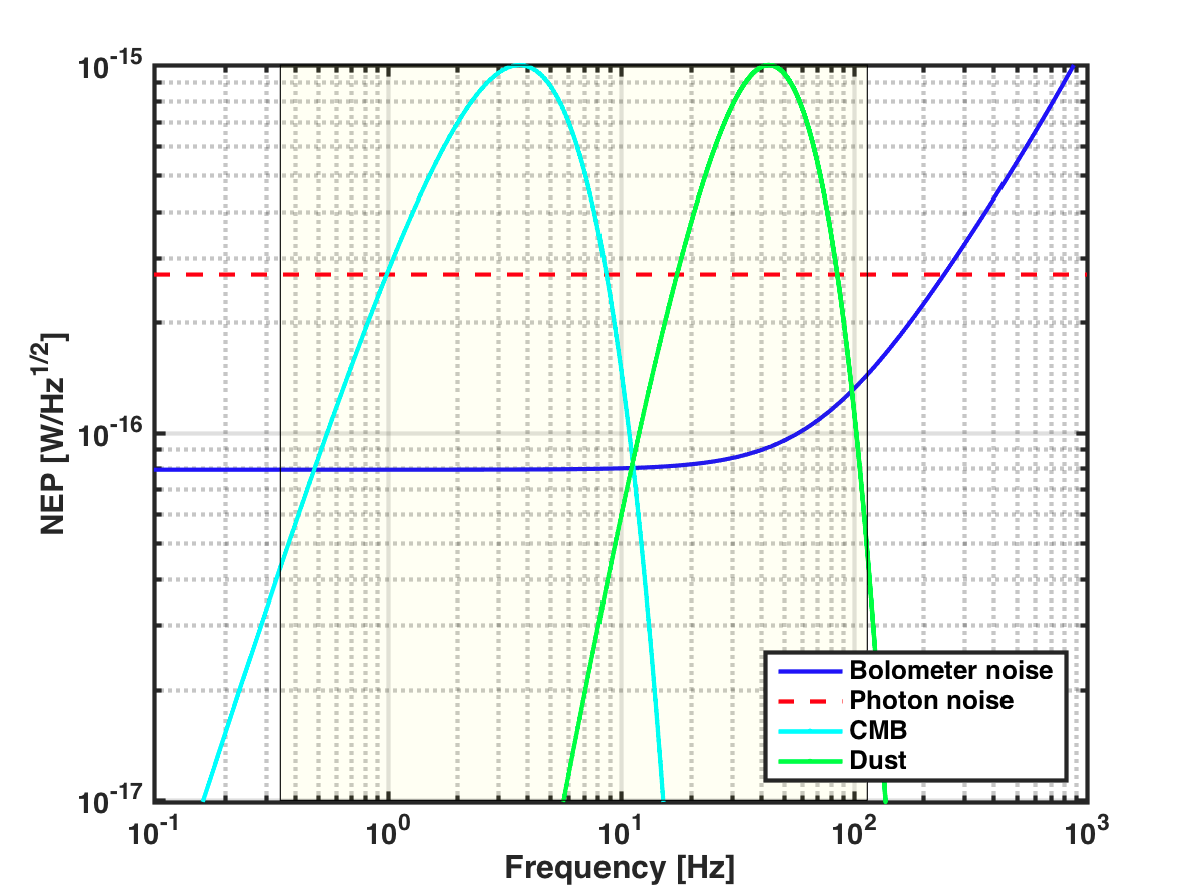}
\caption{Predicted bolometer NEP calculated for the optical and electrical bias conditions in expected in flight. The shaded region shows the PIXIE FTS band. The instrument is photon noise limited across the entire FTS band. Also plotted (not to scale) are the CMB and dust spectra to illustrate how radio frequencies from the sky map to audio frequencies in the PIXIE FTS. The values for the dust temperature and spectral index are based on results from the Planck experiment\cite{Planck2014_b}.}
\label{fig:NEP}
\end{figure}

With relevant parameters taken from load curve analysis (Section \ref{sec:loadcurves}), and accounting for non-equilibrium effects, we calculated the expected bolometer noise under a few bias conditions, and compared them to measured noise spectra. An example noise spectrum and fit is shown in Figure \ref{fig:Se_noise}. Fits are consistently good for multiple measured bias conditions. Running the model for the bias conditions expected during flight, we expect to be photon noise limited across the entire PIXIE bandwidth (Figure \ref{fig:NEP}). 

\section{CONCLUSION}

We designed, fabricated, and characterized large area polarization-sensitive bolometers for the PIXIE experiment. Multimode bolometers designed for a FTS like PIXIE require a different optimization from bolometers developed for focal-plane imagers. In particular, they must handle a large, near-constant optical bias ($\sim 120$ pW), but still operate with sensitivity near the thermodynamic limit across the relevant bandwidth. They also must have a large and mechanically robust absorbing area that is sensitive to all optical frequencies of interest, but relatively insensitive to particle hits.

Mechanical characterization of the fabricated PIXIE bolometers with scanning electron microscopy and white light interferometry shows that our tensioning scheme successfully flattens the absorber strings, enabling indium bump hybridization of a pair of bolometer chips. The resonant frequencies of the tensioned absorbers are also expected to be well above the vibrational frequencies the detectors will experience at launch. We will confirm this in the near future by subjecting a hybridized pair of bolometers to environmental testing.

The dark data provide significant insight into the performance of the PIXIE bolometers, particularly regarding the thermal behavior of the endbanks. Our thermal model agress well with the data, and the results indicate that the PIXIE bolometers satisfy the sensitivity and bandwidth requirements of the space mission (see Figure \ref{fig:NEP}). In the immediate future, we will further characterize the bolometers, specifically targeting the absorber structure. These measurements will include thermal transport measurements of absorbers in the dark, AC impedance measurements, and optical measurements with a cryogenic blackbody source.

\acknowledgments 
 
This work was supported by NASA/GSFC IRAD funding. We are especially grateful to the x-ray microcalorimeter group at NASA/GSFC for lending the Astro-E2/Suzaku test platform for PIXIE detector characterization.



\end{document}